# High efficient sunlight-driven $CO_2$ hydrogenation to methanol over NiZn intermetallic catalysts under atmospheric pressure

Linjia Han[1,2,†], Fanqi Meng[3,†], Xianhua Bai[1], Qixuan Wu[4], Yanhong Luo[1,2,5]*, Jiangjian Shi[1], Yaguang Li[4,*], Dongmei Li[1,2,5], Qingbo Meng[1,5,6,*]

[1]Beijing National Laboratory for Condensed Matter Physics, Institute of Physics, Chinese Academy of Sciences, Beijing, 100190, China.

[2]School of Physical Sciences, University of Chinese Academy of Sciences, Beijing, 100049, P. R. China.

[3]School of Materials Science and Engineering, Peking University, Beijing, 100871, China.

[4]Research Center for Solar Driven Carbon Neutrality, Engineering Research Center of Zero-carbon Energy Buildings and Measurement Techniques, Ministry of Education, The College of Physics Science and Technology, Institute of Life Science and Green Development, Hebei University, Baoding, 071002, China.

[5]Songshan Lake Materials Laboratory, Dongguan, 523808, P. R. China.

[6]Center of Materials Science and Optoelectronics Engineering, University of Chinese Academy of Sciences, Beijing, 100049, P. R. China.

*Corresponding author. Email: qbmeng@iphy.ac.cn; yhluo@iphy.ac.cn; liyaguang@hbu.edu.cn.

†These authors contributed equally to this work.

## Abstract

The synthesis of solar methanol through direct $CO_2$ hydrogenation using solar energy is of great importance in advancing a sustainable energy economy. In this study, non-precious NiZn intermetallic/ZnO catalyst is reported to catalyze the hydrogenation of $CO_2$ to methanol using sunlight irradiation ($\leq 1$sun). The NiZn-ZnO interface is identified as the active site to stabilize the key intermediates of HxCO*. At ambient pressure, the NiZn-ZnO catalyst demonstrates a methanol production rate of 127.5 $\mu$mol $g^{-1} \cdot h^{-1}$ from solar driven $CO_2$ hydrogenation, with a remarkable 100% selectivity towards methanol in the total organic products. Notably, this production rate stands as the highest record for photothermic $CO_2$ hydrogenation to methanol in continuous-flow reactors with sunlight as the only requisite energy input. This discovery not only paves the way for the development of novel catalysts for $CO_2$ hydrogenation to methanol but also marks a significant stride towards a full solar-driven chemical energy storage.

## Introduction

Methanol ($CH_3OH$) is an essential chemical industrial feedstock and also considered as a promising candidate for clean and renewable energy carriers due to its high volume-specific energy density[1, 2]. Currently, the commercial production of methanol is based on syngas as a raw material, requiring synthesis under high temperature and high pressure conditions (3-10 MPa)[3, 4]. This method demands huge energy input and results in additional $CO_2$ emissions. Therefore, the strategy of sunlight driven photo-thermal catalytic hydrogenation of $CO_2$ to produce solar methanol has garnered attention, as this approach can enable safe and efficient storage of solar energy while reducing excess greenhouse gas emissions[5, 6]. Over the past decade, many efforts have been taken to realize efficient photothermic catalytic $CO_2$ hydrogenation to methanol[7-12]. However, these works primarily involve catalysts that contain precious metals, such as ruthenium[7], Indium[8-10, 13, 14], gallium[15, 16], palladium[17], etc. as shown in Supplementary Table 1 and 2, increasing the cost of artificially synthesizing solar methanol. Moreover, the photo-thermal catalytic process mentioned requires intense light irradiation (approximately 6 kW m$^{-2}$ = 6 suns) to generate hot carriers and needs external thermal source for high reaction temperatures (>200 °C)[7, 9]. This leads to additional energy consumption and cost, making the direct use of natural sunlight inconvenient. Therefore, developing economical and highly active catalysts that can efficiently produce methanol from $CO_2$ photo-thermal hydrogenation under atmospheric pressure and natural sunlight (≤1 sun) without additional energy input holds significant scientific and economic value.

In this study, we discovered a NiZn intermetallic/ZnO catalyst (referred to as NiZn-ZnO catalyst) based on non-precious metals that can effectively catalyze the hydrogenation of $CO_2$ to solar methanol under atmospheric pressure. The formation of the NiZn(101)/ZnO(100) heterojunction in the catalyst helps reduce the activation energy and stabilize the key intermediates $H_xCO^*$ during the $CO_2$ hydrogenation process. Under ambient pressure and natural sunlight irradiation (≤1 sun) without with external electric heating, the NiZn-ZnO catalyst achieved a methanol production rate of 127.5 μmol g$^{-1}$ h$^{-1}$, particularly with methanol selectivity reaching 100%. This rate sets a new record for the highest rate of $CO_2$ hydrogenation to methanol driven by natural sunlight in a continuous flow reactor at atmospheric pressure.

Additionally, the NiZn-ZnO catalyst can be synthesized through a conventional co-precipitation method, showing its potential for future large-scale applications. Further, this photo-thermal catalytic method for synthesizing methanol can conveniently integrate with photovoltaic water electrolysis, enabling the production of solar methanol from $CO_2$ and water under full natural sunlight ($\leq$1 sun)[18,19]. This holds significant importance for future large-scale solar energy storage and reducing $CO_2$ emissions.

**Photothermal catalytic performance of NxZy**

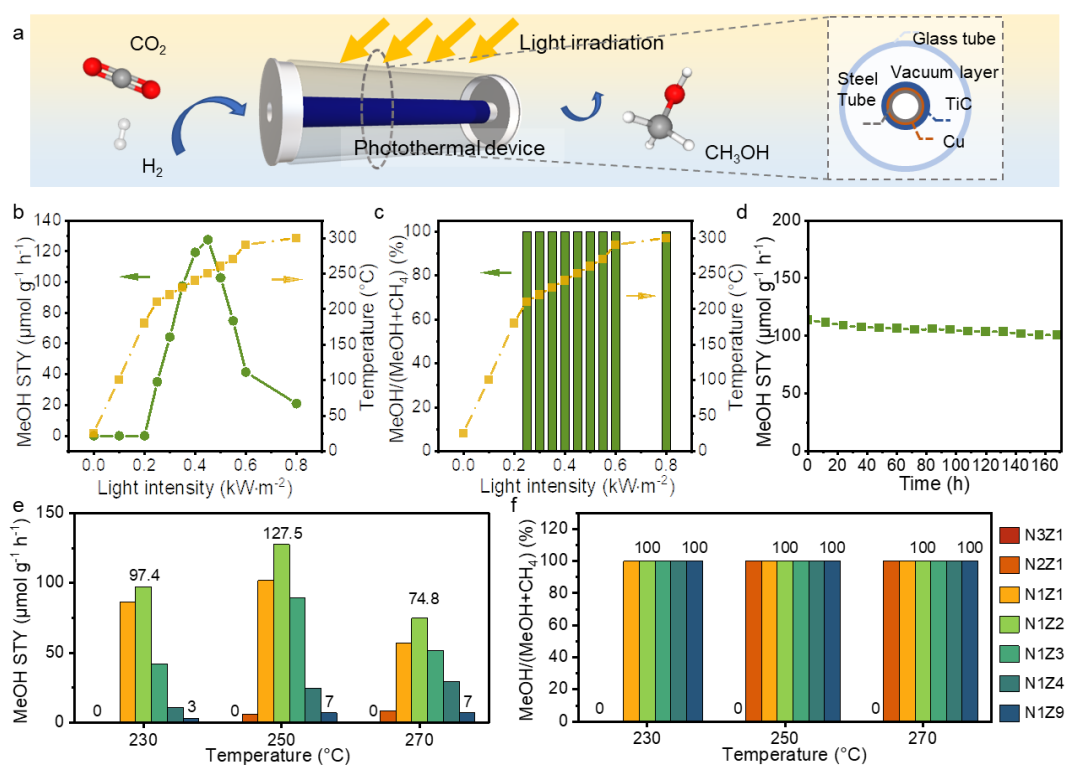

**Fig. 1. The catalytic performance of NxZy catalysts.** (a) The schematic diagram of the photothermal device for $CO_2$ hydrogenation to produce methanol. (b) The temperature and the MeOH space time yield (STY) of N1Z2 catalyst using photothermal device under different sunlight irradiation. (c) The temperature and the selectivity of methanol of N1Z2 catalyst using photothermal device under different sunlight irradiation conditions. (d) The stability test of the sample N1Z2 for more than 5 days at 250 °C. (e) The methanol production rate acquired at 230, 250 and 270 °C using the solar heating device, and (f) the corresponding selectivity of methanol in methanol and methane of NxZy catalysts with different Ni/Zn molar ratios.

A series of NiZn-ZnO catalysts with different Ni/Zn molar ratios were synthesized through a co-precipitation method using metal nitrates as the metal precursor and NaOH solution as the precipitate[20]. The hydroxide precursor was calcined in airflow at 800 °C and subsequently reduced using pure $H_2$ at 500 °C. The obtained catalysts are denoted as NxZy according to the nominal Ni/Zn molar ratio. For example, the sample N1Z2 denotes the catalyst with a Ni/Zn molar ratio of 1:2. The elemental composition of the NxZy catalysts was performed by the X-ray fluorescence technique. As shown in Supplementary Fig. 1, the tested composition is close to the nominal material input. The synthesized NxZy catalysts were then utilized for solar-driven photothermal catalytic $CO_2$ hydrogenation at atmospheric pressure. The schematic diagram of our photothermel device is displayed in Fig. 1a. It is constructed by an outer glass tube and an inner steel tube coated with a TiC layer[21] and a Cu layer on top as shown in the cross-section image in Fig. 1a. These dual-layered coatings act as a light absorber, enabling the absorption of incident light from 300 nm to 1300 nm, which accounts for about 100% absorption of the solar spectrum[21-23]. Between the two tubes, gas is evacuated to form a vacuum layer, preventing the loss of heat. Modulating irradiation from 0 to 1 sun, which is in the intensity range of natural sunlight, as shown in Fig. 1b, this device can heat the loaded catalyst from room temperature to up to 320 °C. This device maintains a typical configuration of a continuous flow fixed bed reactor, and the catalyst is loaded in the innermost tube where reactant gas ($CO_2$, $H_2$) flows through. The $CO_2$ hydrogenation reactions were performed over NxZy catalysts with an $H_2/CO_2$ ratio of 5:1 and a rate of 24 mL min$^{-1}$. Green plots in Fig. 1 b, c display the change of catalytic performance for catalyst N1Z2 along with the increase of the light irradiation intensity, while gold plots in these two figures display the temperature change caused by variation in sunlight irradiation. Methanol generation starts from 180 °C, 0.2 sun, reaches its highest rate of 127.5 μmol g$^{-1}$ h$^{-1}$ at 250 °C, ~ 0.45 sun, and then decreases. Here, we only consider the selectivity towards organic compounds ($CH_3OH$ and $CH_4$), i.e. not considering CO produced by the reverse water–gas shift (RWGS) reaction, and the total conversion ratio of $CO_2$ is given in Supplementary Fig. 2. It is clearly shown in Fig. 1c that no $CH_4$ production is detected till 300 °C, 0.8 sun. Moreover, as shown in Fig. 1d, the methanol production rate by N1Z2 displayed no evident decay for at least 5 days at 250 °C (0.45 sun), indicating the robustness of this catalyst. Fig 1e and 1f demonstrate the performance of the NxZy catalysts (x:y = 1:9 to 3:1). Fig 1e shows the methanol production rate and Fig 1f shows the methanol selectivity in methanol and methane at 230 (0.35 sun), 250 (0.45 sun), and 270°C (0.55 sun). No $CH_4$ production is observed for all these catalysts, which is rarely seen in catalysts with such high Ni content [24]. Among them, at all three temperatures, catalyst N1Z2 exhibited the highest methanol production rate. The performance of the N1Z2 catalyst was also compared with that of monometallic Ni and pure ZnO prepared via the same precipitation procedure, and the results were displayed in Supplementary Fig. 3. The pure ZnO catalyst shows no organic C1 products while $CH_4$ was found only in pure Ni catalysis system. In a previous publication, Zn is shown to have the capacity to modulate Ni's d band energy thus leading to the production of CO instead of $CH_4$ during the $CO_2$ hydrogenation reaction[25]. Herein, the absence of $CH_4$ in the hydrogenation products is in accordance with the result of that study. Moreover, our study here further reveals that another possible strong synergistic effect between Ni and Zn atoms exists, which facilitates the formation of methanol beyond CO during $CO_2$ hydrogenation. Notably, this specific synergistic effect occurs when the Ni/Zn ratios in the NxZy catalysts reach a certain range of values, informing us to focus on the structural differences caused by the composition variation.

**Structure characterization**

To gain structural information and explore the origin of the uncommon catalytic response for NxZy catalysts, a series of physicochemical analyses has been employed. The crystallinity and phase structure of the catalysts were first analyzed by X-ray diffraction (XRD). We assumed that a new active phase was formed during the heat treatment of the catalysts because neither metallic $Ni^0$ nor ZnO were active in the formation of the methanol. Fig 2a displays the XRD patterns of the NxZy catalysts. All the NxZy catalysts exhibit characteristic reflection peaks at 31.81°, 34.5°, 36.3°, and 47.6° of wurtzite ZnO (JCPDF 47-1049), even though the peak intensity becomes very weak for the samples of N1Z1, N2Z1 and N3Z1 with high Ni content. For sample NxZy with Ni/Zn = 1:1 to 1:9, peaks at 43.5° and 46.8° are intensified with increasing amounts of Ni loading, but their peak positions remain almost unchanged. This indicates the formation of NiZn intermetallic (JCPDF 06-0672) with equimolar Ni and Zn. In the samples with Ni/Zn ratio higher than 1 (N2Z1 and N3Z1), new peaks at 43.9° and 51.1° near those of pure metallic Ni[0] appear but shift toward lower 2θ angles, suggesting the formation of Ni-rich Ni-Zn alloy phases. The emergence of these structures is in consistency with previous publications, that high-temperature heat treatment will induce diffusion between Zn and Ni atoms, thus leading to the formation of NiZn intermetallic and/or Ni-rich Ni-Zn alloy phases. [20,26,27]

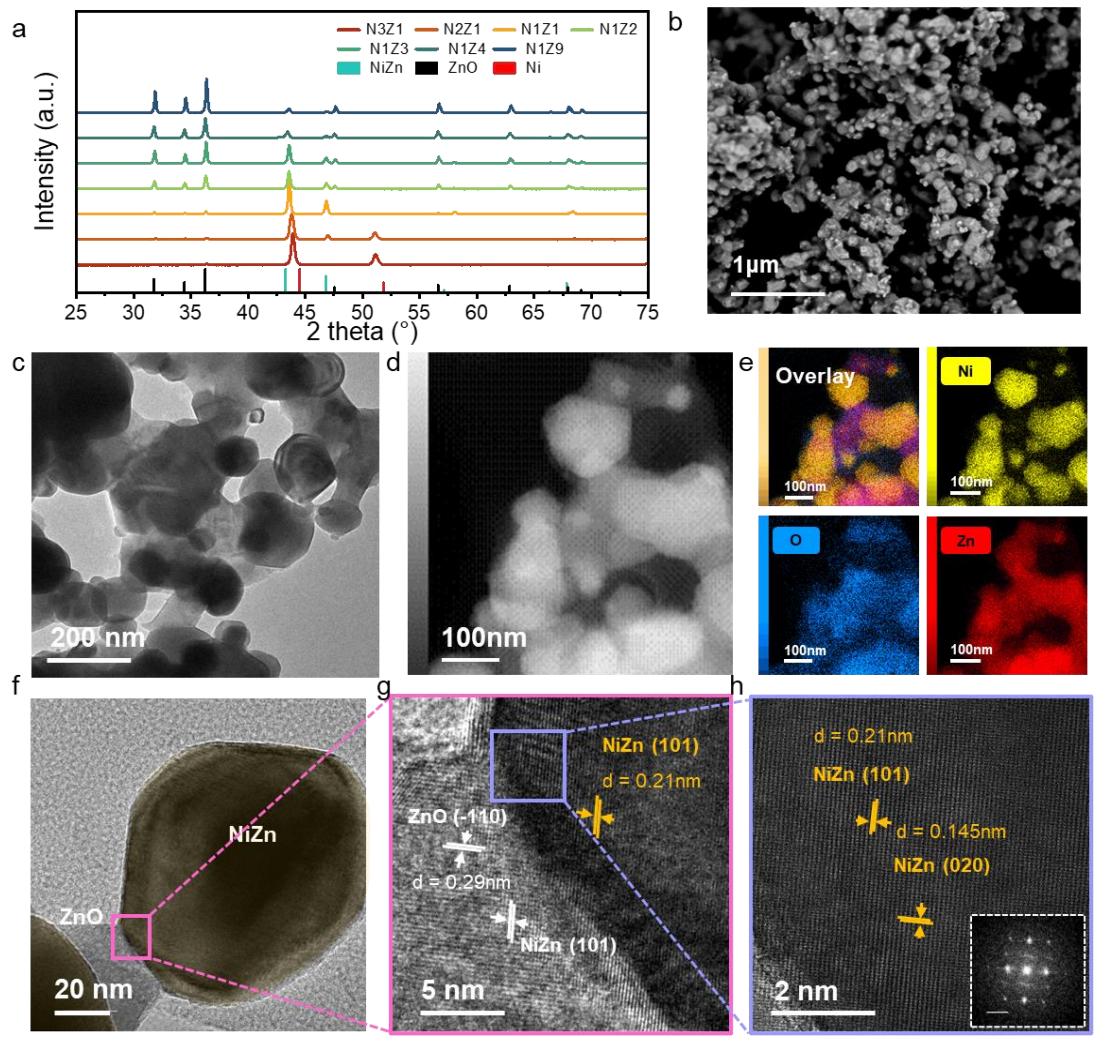

**Fig. 2: Structural characterization of the N1Z2 catalyst.** (a) The XRD patterns of NxZy catalysts, and (b) the SEM image, (c) the TEM image, (d-e) the HAADF-STEM-EDS image, and (f-h) the representative HRTEM images of the N1Z2 catalyst, the picture inset in (h) is fast Fourier transform image and the scale bar is 0.5 1/nm.

Combining the catalytic performance of NxZy with the XRD data, it is shown that only catalysts with both the structure NiZn intermetallic and ZnO can produce methanol. We then analyzed the morphology of the best-performing catalyst N1Z2 using scanning electron microscope (SEM) and transmission electron microscope (TEM), and the images are presented in Fig. 2b and 2c. The SEM and TEM images demonstrate an aggregated structure of small regular nanoparticles dispersed by a thin plate-like phase. The high-angle annular dark-field scanning transmission electron microscopy (HAADF-STEM) image (Fig. 2d) and elemental mapping (Fig. 2e) reveal that the Zn is well distributed throughout the whole test area, while O is predominantly distributed throughout the plate-like domain but is deficient in the regular particles where element Ni prevails. This suggests that the observed nanoparticle is NiZn intermetallic, while the plate-like phase is ZnO. The interface structure between NiZn intermetallic and ZnO is further investigated in detail. The high-resolution TEM (HRTEM) images of the N1Z2 catalyst are shown in Fig. 2f-h. The corresponding fast Fourier transform (FFT) image of the NiZn intermetallic area is shown inset Fig. 2h and those of the areas close to the interface are shown in Supplementary Fig. 4. Fig. 2g presents a magnified view of the NiZn/ZnO interface in N1Z2 catalyst depicted in Fig 2f. Two different lattice fringe spacing were observed at the nanoparticle and plate-like phase sides respectively. One is 0.21nm, corresponding to NiZn intermetallic (101), and the other is 0.29 nm, corresponding to ZnO (100). Notably, close to the interface, the d-spacing of 0.21 nm is also observed on the side of ZnO. This indicates that a thin layer of NiZn forms on top of ZnO or Ni atoms diffuse into the ZnO lattice. Fig. 2h shows an HRTEM image of the NiZn intermetallic particle. The inset picture demonstrates the FFT image of this area. This diffraction pattern matches well with the pattern observed along NiZn[101] orientation and is consistent with the XRD pattern of the catalyst. In addition, the relative intensity of NiZn (101) is higher than the standard pattern, indicating its tendency to stack perpendicularly to this plane, whereas ZnO does not exhibit a preferential orientation[28].

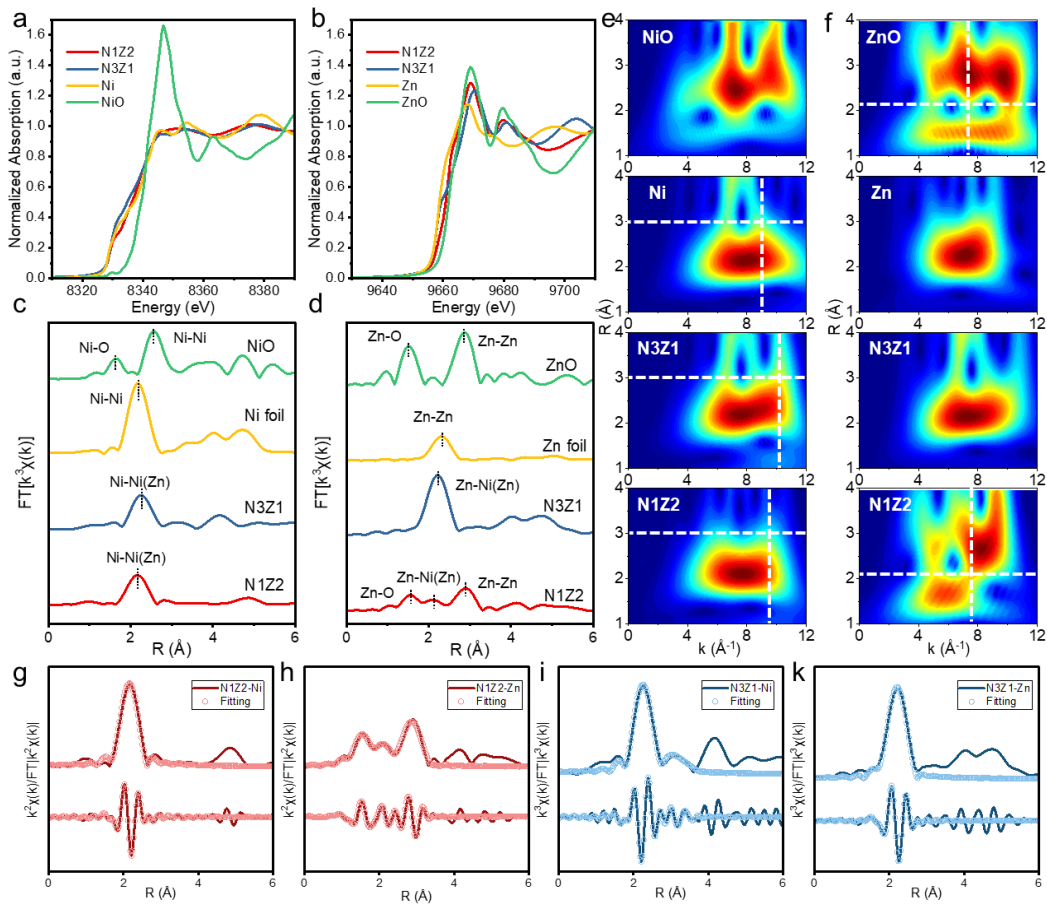

**Fig. 3 Electronic property and local structure of the catalysts. (**a) Ni K-edge and (b) Zn K-edge XANES spectra of N1Z2, N3Z1, and reference samples. (c) Ni and (d) Zn $k^3$-weighted FT of EXAFS spectra of these samples. (e) Ni K-edge and (f) Zn K-edge WT–EXAFS spectra of these samples. (g, h) The first two shells: (g) Ni-Zn/Ni-Ni and (h) Zn-O, Zn-Ni, and Zn-Zn fittings of the FT–EXAFS spectra for N1Z2. (i, k) Fitting curves of the FT–EXAFS spectra for N3Z1.

To gain a thorough understanding of the local structures and coordination environment of the N1Z2 catalyst, X-ray absorption fine structure (XAFS) spectroscopy was conducted. We use the N3Z1 catalyst with neither the NiZn intermetallic phase nor the catalytic capacity to produce methanol as the control group for comparison. As shown in Fig. 3a, the Ni X-ray absorption near-edge structure (XANES) edge of both N1Z2 and N3Z1 is closer to that of Ni foil instead of NiO, indicating that the valance state of Ni is close to 0. The Zn K-edge XANES spectra in Fig. 3b show that the absorption edge of sample N1Z2 and sample N3Z1 lie between those of Zn foil and ZnO, suggesting that the average valance state of Zn is between 0 and $+2^{29}$. Moreover, the edge of N3Z1 is closer to that of Zn foil compared to N1Z2. This result aligns well with the XRD data, which demonstrates that the sample N3Z1 maintains a bulk structure like that of metallic $Ni^0$ while sample N1Z2 preserves a composite structure of ZnO and metallic $Zn^0$. Fig. 3c displays the Ni K-edge Fourier transform extended XAFS (FT-EXAFS) spectra of the samples N1Z2 and N3Z1. The spectra of both the catalysts are like that of Ni foil: a peak at 2.18 Å and 2.24 Å appeared in the spectrum of N1Z2 and N3Z1, respectively, indicating the presence of Ni or Zn in the nearest Ni neighbors. In Fig. 3d, the Zn K-edge EXAFS spectra of N3Z1 show a single main peak centered at 2.21 Å, which can be attributed to the Zn-Ni(Zn) bond. For sample N1Z2, two main peaks centered at 1.50 and 2.88 Å can be attributed to Zn-O and Zn-Zn bonds in ZnO phase, while the lower peak in the middle can be attributed to Zn-Ni(Zn) bond. Due to the similarity between Ni-Ni and Ni-Zn bonding, it is hard to distinguish one from another in the FT-EXAFS spectra[20,30,31]. To further clarify this electronic interaction, wavelet transform (WT)-EXAFS and DFT-based optimizations were conducted to probe the adjacent properties of the metal atoms. The (WT)-EXAFS spectra of both Ni and Zn are in good agreement with the FT results. In Fig. 3e, neither in the spectrum of N1Z2 nor in that of N3Z1 can observe Ni-O scattering in 1.65 Å. In Fig 3f, the spectrum of N1Z2 is more like that of ZnO while that of N3Z1 is more like that of Zn foil. Since the backscattering factor F(k) strongly depends on the atomic number Z, heavy atoms will contribute more at higher wavenumbers in the (WT)-EXAFS spectrum compared to the lighter ones[31]. As shown in Fig. 3e, compared to the spectrum of Ni foil, the spectra of N1Z2 and N3Z1 at 3 Å display maximums at higher k, indicating the presence of Ni-Zn scattering. It is also worth noting that in the spectrum of N1Z2, signals associated with Ni-Ni scattering in the high R region are absent, suggesting that Zn atoms are in the neighboring sites of Ni atoms. In the Zn (WT)-EXAFS spectra shown in Fig 3f, the spectrum of N1Z2 displays magnified intensity at 2.2 Å, which provides further evidence of the presence of Ni-Zn scattering. As for the spectrum of N3Z1, there is maximum present at higher R, indicating the presence of Ni atom in the neighbor of Zn atoms. The fitting results of the EXAFS spectra are shown in Fig. 3g and 3i in R space. According to the bulk structure, the plot of N3Z1 is fitted with the first two neighbor shells of Ni (Fig. 3i, k) plus the scattering path of Ni-Zn, and the signal of sample N1Z2 is fitted with the scattering path from both NiZn and ZnO. The main peak in Fig. 3g is constructed by the first two neighbor shells of NiZn (Ni-Zn and Ni-Ni), and the plot in Fig. 3h is fitted by the Zn-O path and Zn-Zn path in ZnO and the Ni-Zn path in NiZn. The obtained EXAFS fitting curves in R-space and k-space are consistent with both the Ni and Zn experimental spectra of both samples. The coordination numbers (CNs), bond distances (R), and Debye–Waller factor ($\sigma^2$) are listed in Supplementary Table 3-4. The existence of Ni-Zn scattering further implies the strong interaction between Ni and Zn, shedding light on the elucidation of $CO_2$ to methanol mechanism in the NxZy catalysts.

**Determination of active sites.**

In all the NxZy catalysts, only samples containing both ZnO and NiZn intermetallic phases in their bulk structures display the ability to produce methanol. From the EXAFS results, compared to N3Z1, the spectrum of N1Z2 with the highest methanol production rate displays absent signals related to Ni-Ni scattering in the higher R region. It is reasonable to deduct that the active phase should be correlated with NiZn intermetallic. To further determine the origin of the methanol formation capability and avoid the influence of Ni doping for ZnO, ZnO doped with a minute amount of Ni (Ni: Zn = 3: 97, denoted as Ni-ZnO) was synthesized, and the XRD pattern is shown in Supplementary Fig. 5. Compared with those of pure ZnO, the peak positions of Ni-ZnO slightly shift towards higher angles, indicating the successful doping of Ni into the ZnO lattice. Under the same catalytic reaction condition, as shown in Supplementary Fig. 3, the Ni-ZnO sample shows neither $CH_4$ nor MeOH production. Therefore, its effect on methanol production can be excluded. Supplementary Fig. 6 provides the HAADF-STEM-EDS images of N1Z1. It is clear for sample N1Z1, the NiZn intermetallic is in predominance both in bulk and at the surface, while a much smaller amount of ZnO is found compared to that in sample N1Z2. Curiously, sample N1Z1 displayed a comparable level of methanol production with sample N1Z2 despite their structural differences (Fig. 1e). Since ZnO can facilitate $CO_2$ adsorption thus fostering methanol production, combined with the surface structure discussed earlier, we finally postulate that the active phase for methanol formation is the interface of NiZn/ZnO.

**The formation mechanism of methanol in NxZy**

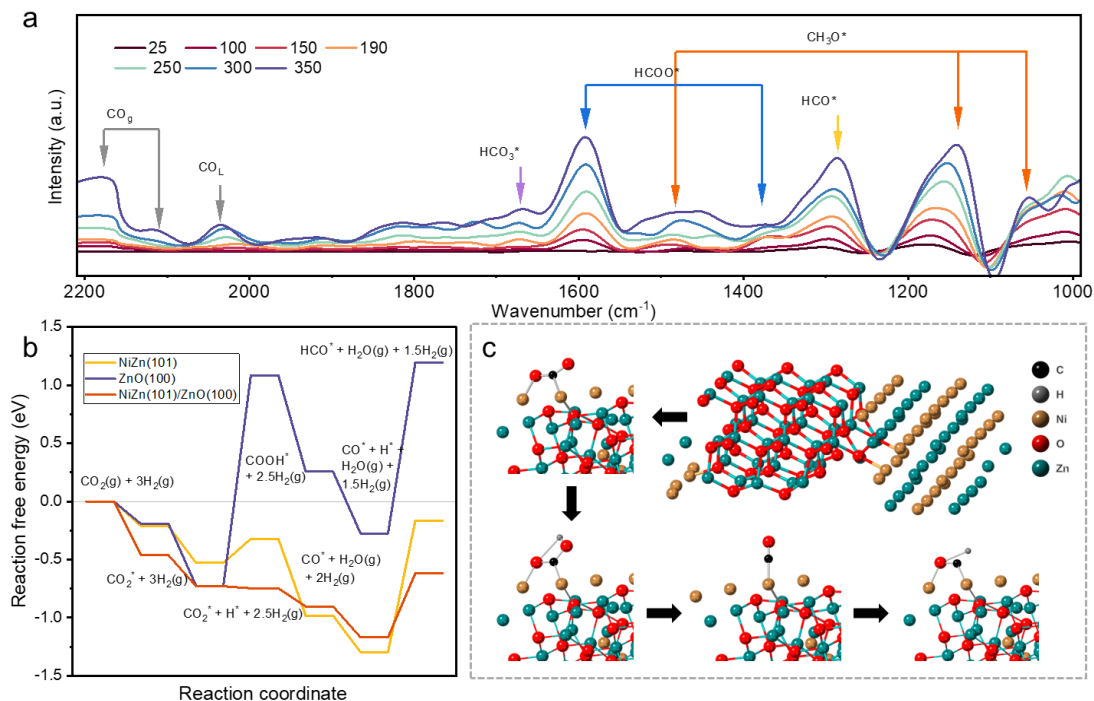

**Fig. 4 Mechanistic studies of the catalytic activity of NiZn-ZnO in $CO_2$ hydrogenation.** (a) The DRIFTS spectra of fresh N1Z2. (b) The potential energy diagrams obtained through DFT calculations for the catalytic $CO_2(g)$ hydrogenation to $CH_3OH(g)$ on NiZn(101), ZnO(100), and their heterojunction (NiZn(101)/ZnO(100)). (c) Proposed reaction mechanism for the conversion of $CO_2$ to methanol over NiZn-ZnO.

To delve into the reaction pathway of $CO_2$ hydrogenation to methanol, we undertook in-situ Diffuse Reflectance Infrared Fourier Transform Spectroscopy (in-situ DRIFTS) studies. These investigations were carried out at temperatures spanning from room temperature to 400 °C on sample N1Z2. In the spectra of the freshly reduced sample N1Z2 (Fig. 4a), several surface species were detected (Supplementary Table 5), including formate species (HCOO*) at 1591 $cm^{-1}$ and 1371 $cm^{-1}$ [24,32], carbonate species ($HCO_3$*) at 1660 $cm^{-1}$ and 1690 $cm^{-1}$, and methoxy species (HCO*, $CH_3O$) at 1065, 1140, 1289, and 1480 $cm^{-1}$ [33,34] Additionally, gas-phase CO ($CO_g$) is observed at 2100 cm$^{-1}$ and 2170 cm$^{-1}$, and linearly adsorbed CO ($CO_l$) is present at 2030 cm$^{-1}$. These findings suggest the existence of two plausible reaction pathways for the hydrogenation of $CO_2$ to methanol: one involves the formation of formate intermediates, known as the formate pathway, while the other requires the initial cleavage of the C=O bond, resulting in the creation of chemisorbed CO, followed by its hydrogenation to produce methanol[35]. To elucidate the mechanism of methanol formation, we compare the DRIFTS spectra of fresh N2Z1 and deactivated N2Z1 (Supplementary Fig. 7), with a primary emphasis on discerning the variations in surface species. At 250°C (green line), both spectra exhibit peaks belonging to HCOO$^*$ at nearly identical positions with comparable signal intensity. However, concerning peaks associated with $CH_3O^*$ and HCO$^*$, a substantial disparity in peak positions is evident between these two spectra. Compared to the fresh catalyst, peaks related to $CH_3O^*$ and HCO$^*$ in the deactivated catalyst downshift from 1140 to 1124.7 cm$^{-1}$ and from 1289 to 1254.5 cm$^{-1}$ respectively. It appears that the HCOO$^*$ species is not highly sensitive to the variations in surface states, while $CH_3O^*$ and HCO$^*$ are significantly influenced. Additionally, temperature serves as another influential factor in determining the peak positions of $CH_3O^*$ and HCO$^*$. To investigate the changes in the surface chemical state induced by the catalytic reaction, we performed semi-in-situ X-ray Photoelectron Spectroscopy (semi-in-situ XPS) on the sample N1Z2 at stages where it was freshly reduced and after a long time of reaction (denoted as fresh and deactivated in Supplementary Fig. 8). All peaks presented were calibrated by the C 1s spectrum at 284.8 eV. In the Zn 2p spectra of the freshly reduced catalyst, two sets of peaks are observed. Taking peaks of Zn 2p$_{3/2}$ as an example, the primary peak is located at 1020.9 eV, accompanied by a smaller peak at 1021.5 eV, indicating the existence of two kinds of Zn with different valence states. After a long time of reaction, only one set of peaks can be found, and the main peak of Zn 2p$_{3/2}$ shifts from 1020.9 to 1021.3 eV. The enhanced binding energy shows that surface Zn is oxidized. Due to the subtle difference in binding energies between $Zn^{2+}$ and $Zn^0$ (ca. 0.3 eV), we cannot tell the exact chemical state of $Zn^0$ in the N1Z2 catalyst. In the Ni 2p spectra, a similar trend is observed, with an increase in binding energies for both $Ni^0$ from 851.7 to 852.1 eV and $Ni^{2+}$ from 854.5 to 855.1 eV, indicating the oxidation of Ni. In O 1s spectra, an additional peak at 532.2 eV emerges after the reaction, while peaks associated with lattice oxygen exhibit no conspicuous changes in binding energy. The new peak can be attributed to either oxygen vacancy or surface-adsorbed hydroxyl groups[36]. Previous research on $CO_2$ hydrogenation to methanol via supported Cu catalysts has shown that, in certain instances, HCOO[*] species may not function as a reaction intermediate but rather as a spectator, due to its strong bonding to the surface[35]. This insight illuminates that the nearly identical signal of HCOO[*] is detected in the DRIFTS spectra of both the fresh and deactivated catalysts in our research.

Summing up the above discussion, we propose that methanol is formed through C-O bond cleavage and subsequent CO hydrogenation, at the interface of NiZn and ZnO. The results of Density Functional Theory (DFT) calculations further support this hypothesis. The calculated potential energy diagrams of the catalytic $CO_2$ (g) hydrogenation to $CH_3OH$(g) on NiZn (101), ZnO (100), and their heterojunction (NiZn (101)/ZnO (100)) are shown in Fig 4b. The chemisorption of $CO_2$ on each surface can be easily achieved since the activation energy of the first two steps is negative. However, hydrogenation can merely happen in ZnO (100) due to the high activation energy. This can be explained by the lack of active metal for the dissociation of $H_2$ molecules in ZnO. Compared to ZnO (100), NiZn (101) displayed a sharp decrease in active energy at the same step of chemisorbed $CO_2$ hydrogenating to COOH[*] with carbon as the adsorption atom. The formation of heterojunction of NiZn and ZnO further lowers the energy required in the hydrogenation step, enhancing methanol production under the same reaction condition. The other hydrogenation step of linear adsorbed CO adding one H atom to form the HCO[*] intermediate follows the same trend of decrease in the activation energy in the formation of NiZn (101)/ZnO (100). The mechanism for methanol formation is proposed and illustrated in Fig 4c. $CO_2$ is firstly chemisorbed on the side of NiZn (101) in a bent configuration, then the addition of one H atom on top of it leads to the formation of COOH[*] and subsequent cleavage of the C-O bond, forming linear adsorbed CO. Subsequent hydrogenation of $CO_1$ via HCO[*] intermediate finally results in the formation of methanol. In the DRIFTS spectra of the ZnO sample and N1Z2 sample shown in Supplementary Fig. 9 intermediates related to CO hydrogenation ($CH_3O$[*] and HCO[*]) can only be found in that of the N1Z2 sample while formate species are found in all the spectra. This difference provides additional evidence to support that in the N1Z2 catalyst, methanol is formed through the CO hydrogenation pathway.

**Conclusion**

In summary, we demonstrate a novel NiZn intermetallic/ZnO catalyst for excellent methanol generation from $CO_2$ hydrogenation at ambient pressure. This NiZn-ZnO catalyst realized a solar methanol generation rate of 127.5 µmol $g^{-1}$ $h^{-1}$ with zero $CH_4$ generation as a byproduct at atmospheric pressure and under sunlight irradiation without external thermal source. The conjunction of NiZn intermetallic and ZnO phases is the key for methanol production during $CO_2$ hydrogenation. DRIFTS and semi-in-situ XPS studies identified that the reaction pathway of $CO_2$ hydrogenation to methanol involves the initial decomposition of $CO_2$ followed by CO hydrogenation. Potential energy diagrams obtained through DFT calculations reveal that the formation of the NiZn(101)/ZnO(100) heterojunction helps lower the activation energy in hydrogenation steps, stabilizing key intermediates of CO hydrogenation ($CH_3O*$ and $HCO*$), thus achieving methanol production. The methanol production rate was maintained for up to 5 days on-stream during the long-term stability test, which demonstrated the robust nature of the NiZn intermetallic/ZnO catalyst. Therefore, our study provides a rational design of Ni-based catalysts for tuning the $CO_2$ activation and conversion pathways. Furthermore, this photothermal catalytic method for synthesizing methanol can conveniently integrate with the photovoltaic water electrolysis of green $H_2$ production systems, realizing the production of solar methanol from $CO_2$ and water under natural sunlight ($\leq$1 sun). This has significant implications for future large-scale solar energy storage and reducing $CO_2$ emissions.

**Acknowledgements**

The authors gratefully acknowledge the cooperation of the beamline scientists, Dr. Xueqing Xu and Dr. Yunpeng Liu, at SAXS/XRD/XAFS beamline 1W2B of the Beijing Synchrotron Radiation Facility (BSRF-1W2B). This work was supported by the National Natural Science Foundation of China (Grant nos. 52227803 (Q. M.) and 52172261 (Y. L.)). J. S. also gratefully acknowledges the support from the Youth Innovation Promotion Association of the Chinese Academy of Sciences.

**Author contributions**

Linjia Han, Yanhong Luo and Qingbo Meng conceived the idea and designed the experiments. Linjia Han and Fanqi Meng did the experiments and the data analysis. Xianhua Bai, Jiangjian Shi, Qixuan Wu and Dongmei Li supported catalyst characterization, performance test and discussions. Linjia Han, Yanhong Luo, Yaguang Li and Qingbo Meng participated in writing the manuscript.

**Competing interests**

There is no interest conflict.